\documentclass[journal]{IEEEtran}
\usepackage[T1]{fontenc}
\usepackage[latin9]{inputenc}
\usepackage{float}
\usepackage{booktabs}
\usepackage{bm}
\usepackage{amsmath}
\usepackage{amssymb}
\usepackage{graphicx}

\makeatletter

\providecommand{\tabularnewline}{\\}
\floatstyle{ruled}
\newfloat{algorithm}{tbp}{loa}
\providecommand{\algorithmname}{Algorithm}
\floatname{algorithm}{\protect\algorithmname}
\usepackage{tikz}
\usetikzlibrary{calc}

\usepackage[caption=false,font=scriptsize,labelfont=rm,textfont=rm,subrefformat=parens]{subfig}
\usepackage{algorithmic}
\usepackage{diagbox}
\usepackage{cite}

\makeatother

\begin{document}
\title{Meta-Reinforcement Learning With Mixture of Experts for Generalizable
Multi Access in Heterogeneous Wireless Networks}
\author{Zhaoyang~Liu, Xijun~Wang,\IEEEmembership{} Chenyuan Feng, Xinghua
Sun, Wen Zhan, and~Xiang~Chen\thanks{Part of this work was presented at the WiOpt 2023 Workshop on Machine
Learning in Wireless Communications \cite{liu.chen2023}.}\thanks{Z.~Liu, X.~Wang, and X.~Chen are with the School of Electronics
and Information Technology, Sun Yat-sen University, Guangzhou, China
(e-mail: liuzhy86@mail2.sysu.edu.cn; wangxijun@mail.sysu.edu.cn; \mbox{chenxiang@mail.sysu.edu.cn}).}\thanks{C. Feng is with the Department of Communication Systems, EURECOM,
Biot, France. (e-mail: Chenyuan.Feng@eurecom.fr).}\thanks{X. Sun and W. Zhan are with the School of Electronics and Communication
Engineering, Shenzhen Campus of Sun Yat-sen University, Shenzhen,
China (e-mail: \mbox{sunxinghua}@mail.sysu.edu.cn; zhanwen6@mail.sysu.edu.cn).}}
\maketitle
\begin{abstract}
This paper focuses on spectrum sharing in heterogeneous wireless networks,
where  nodes with different  Media Access Control (MAC) protocols
to transmit data packets to a common access point over a shared wireless
channel. While previous studies have proposed Deep Reinforcement Learning
(DRL)-based multiple access protocols tailored to specific scenarios,
 these approaches are limited by their inability to generalize across
diverse environments, often requiring time-consuming retraining.
To address this issue, we introduce Generalizable Multiple Access
(GMA), a novel Meta-Reinforcement Learning (meta-RL)-based MAC protocol
designed for rapid adaptation across heterogeneous  network environments.
GMA leverages a context-based meta-RL approach with Mixture of Experts
(MoE) to improve representation learning, enhancing latent information
extraction.  By learning a meta-policy during training, GMA enables
 fast adaptation  to different and previously unknown environments,
without prior knowledge of the specific MAC protocols in use.  Simulation
results demonstrate that, although the GMA protocol experiences a
slight performance drop compared to baseline methods in training environments,
it achieves faster convergence and higher performance in new, unseen
environments.
\end{abstract}


\IEEEpeerreviewmaketitle{}

\thispagestyle{empty}

\section{Introduction}

In contemporary wireless environments, multiple network types with
distinct characteristics, such as 5G, WiFi, and IoT networks, coexist
within the same physical space, each using different Media Access
Control (MAC) protocols. This diversity presents significant challenges
for spectrum sharing and interference management. Traditionally, networks
have relied on pre-allocating exclusive frequency bands, a method
that often leads to inefficient spectrum utilization. As wireless
technologies continue to proliferate, the demand for wireless communication
services is outpacing the availability of spectrum resources. To address
this, the Defense Advanced Research Projects Agency (DARPA) introduced
the Spectrum Collaboration Challenge (SC2) competition \cite{.cy,tilghman.tilghman201906},
proposing a novel spectrum-sharing paradigm. In this approach, co-located,
heterogeneous networks collaborate to share spectrum without requiring
prior knowledge of each other's MAC protocols. SC2 encourages the
development of intelligent systems capable of adapting in real-time
to dynamic and congested spectrum environments. A key solution to
this challenge lies in developing intelligent MAC protocols for a
particular network among those sharing the spectrum, enabling efficient
and equitable spectrum sharing.

Several Deep Reinforcement Learning (DRL) based multiple access schemes
have been developed for coexistence with heterogeneous wireless networks~\cite{yu.liew201906,ye.fu202202,yu.wang202109,xu.quek202205,ye.fu202005,geng.zheng2022,deng.han202205,yu.wang202210,lu.chen202205,han.sun2024}.
One notable protocol in this area is the Deep Reinforcement Learning
Multiple Access (DLMA) proposed by Yu \emph{et al.}~\cite{yu.liew201906},
which utilizes Deep Q-Network (DQN)~\cite{mnih.hassabis201502} to
learn the access policy. Specifically, the DRL agent in DLMA makes
access decisions based on historical data to enable coexistence with
other nodes that use different MAC protocols. The DRL-based multiple
access scheme can effectively learn to coexist with other MAC protocols
in specific pre-determined heterogeneous wireless network scenarios.
However, a key challenge that remains unaddressed is the ability to
generalize DRL performance to unseen testing environments that differ
from the training environments. In real-world deployments, the heterogeneous
wireless network composition, including the number of nodes and MAC
protocols employed, is likely to vary across different environments.
This mismatch between training and testing environments can lead to
degraded performance of the DRL agent. While it is possible to train
a new DRL model from scratch for each unseen testing environment,
this approach is highly inefficient and time-consuming due to the
significant effort required to collect sufficient training data and
the high sample complexity involved in achieving convergence. Therefore,
enhancing the generalization capabilities of DRL agents to maintain
robust performance across diverse, unseen heterogeneous wireless environments
is a crucial  challenge.

In this paper, we investigate harmonious spectrum sharing in co-located
heterogeneous wireless networks, where a DRL-based agent node and
multiple existing nodes with different MAC protocols share the same
wireless channel. Our primary objective is to enhance the generalization
capabilities of DRL-based multiple access control across diverse coexistence
scenarios, enabling rapid adaptation to previously unseen and dynamic
wireless environments.  In our prior work \cite{liu.chen2023}, we
 focused solely on maximizing the total throughput of the heterogeneous
wireless network, without considering fairness  between the agent
node and the existing nodes. We expand this research by addressing
the critical issue of fairness in the co-existence scenario, fostering
diverse and equitable coordination patterns. Additionally, we incorporate
a Mixture of Experts (MoE) architecture into meta Reinforcement Learning
(meta-RL) approach to improve task representation capabilities.  The
key contributions of this paper are summarized as follows:
\begin{itemize}
\item We consider a range of heterogeneous wireless environments, treating
each as a distinct task in the context of meta-RL. In each environment,
we model the multiple access problem in a heterogeneous wireless network
as a Markov Decision Process (MDP). To balance the dual objectives
of maximizing throughput and ensuring fairness, we define a reward
function that considers both system throughput and fairness between
the agent node and existing nodes. We further structure the problem
within a meta-RL framework to ensure the generalizability of the learned
policy across varying network conditions and diverse environments.
\item We propose a novel MAC protocol for the agent node, called Generalizable
Multiple Access (GMA),  which leverages context-based off-policy meta-RL
with MoE layers to enhance the agent node's ability to make intelligent
access decisions across diverse network environments. The GMA protocol
includes an MoE-enhanced encoder to generate more discriminating task
embeddings and uses Soft Actor-Critic (SAC) for learning a goal-conditioned
policy.  Through this approach, GMA learns a meta-policy based on
experiences from a variety of training tasks, rather than task-specific
policies. This meta-policy allows GMA to quickly adapt to new tasks
or environments, significantly improving convergence speed when faced
with previously unseen scenarios.
\item Simulation results demonstrate that GMA ensures universal access in
well-trained environments, delivers high initial performance in new
environments, and rapidly adapts to dynamic conditions. We also show
that, with our proposed fairness metric, GMA effectively balances
high total throughput with fairness.  Through extensive experiments,
we explore the impact of training task selection on zero-shot and
few-shot performance and provide valuable insights into designing
effective meta-learning training sets. Moreover, we highlight the
performance improvements enabled by the MoE architecture, underscoring
the superiority of our approach, and examine how the number of experts
influences system performance.
\end{itemize}

The remainder of this paper is organized as follows. In Section \ref{sec:Related-Work},
we review related work.  Section~\ref{sec:model} details the system
model and problem formulation. In Section~\ref{sec:protocol}, we
present the proposed meta-RL-based MAC protocol. The simulation results
are discussed in Section~\ref{sec:results}. Finally, Section~\ref{sec:Conclusion}
concludes with the main findings.

\section{Related Work\label{sec:Related-Work}}

\subsection{DRL in Heterogeneous Wireless Networks}

The concept of heterogeneous wireless networks, where diverse communication
protocols coexist in the same physical space and share the same spectrum,
has emerged as a promising paradigm to achieve higher spectral efficiency.
A wide range of MAC protocol designs for heterogeneous wireless networks
have drawn inspiration from DLMA \cite{yu.liew201906}, with various
extensions proposed to address different coexistence scenarios. In~\cite{ye.fu202202},
DLMA has been extended to address the multi-channel heterogeneous
network access problem, where the DRL agent decides both whether to
transmit and which channel to access. A variant called CS-DLMA was
introduced in \cite{yu.wang202109}, incorporating  carrier sensing
(CS) capability to enable coexistence with carrier-sense multiple
access with collision avoidance (CSMA/CA) protocols.  The authors
in~\cite{xu.quek202205} further introduced a  MAC protocol enabling
DRL nodes to access the channel without detecting the channel idleness,
and assessed the coexistence performance  with WiFi nodes. A MAC protocol
based on DRL was proposed in~\cite{ye.fu202005} to coexist with
existing nodes in underwater acoustic communication networks, where
high delay in transmissions is a concern. On the basis of~\cite{ye.fu202005},
the work in \cite{geng.zheng2022} further addressed the issue of
coexistence with asynchronous transmission protocol nodes in underwater
acoustic communication networks.  Deng \emph{et al.}~\cite{deng.han202205}
proposed an R-learning-based random access scheme specifically designed
for coexistence in heterogeneous wireless networks with delay-constrained
traffic. For scenarios involving multiple agent nodes, distributed
DLMA was introduced in~\cite{yu.wang202210} to facilitate the coexistence
of multiple agents in heterogeneous wireless networks with imperfect
channels. In~\cite{han.sun2024}, a novel framework leveraging curriculum
learning and multitask reinforcement learning was introduced to enhance
the performance of access protocols in dynamic heterogeneous environments.
Additionally, a QMIX-based multiple access scheme was proposed for
multiple nodes in~\cite{guo.sun202205}, which also demonstrates
compatibility with CSMA/CA protocols.  However, the majority of these
works neglects challenges in unseen and dynamic environments, rendering
the trained policies effective only for scenarios  similar to those
used in training. Only a few studies, such as \cite{ye.fu202202,han.sun2024},
have considered this issue, but they solely rely on recurrent neural
network and require extensive gradient updates for adaptation in dynamic
environments.

\subsection{Fairness Coexistence}

Ensuring fairness among nodes with different protocols is crucial
in coexistence environments.  A common approach is to incorporate
the fairness metric directly into the objective function and redesign
the reward function to account for fairness considerations. In light
of this, DLMA~\cite{yu.liew201906} modified the standard Q-learning
algorithm by incorporating an $\alpha$-fairness factor into the Q-value
estimate to meet the fairness objective. Frommel \emph{et al.}~\cite{frommel.larroca202303}
proposed a DRL approach to dynamically adjust the contention window
of 802.11ax stations. They adopted the $\alpha$-fairness index as
the metric of the overall system performance, rather than sum throughput,
to simultaneously improve raw data rates in WiFi systems and maintain
the fairness between legacy stations and 802.11ax stations. The authors
in~\cite{pei.zhang202304} introduced a mean-field based DRL approach
for coexistence with WiFi access points, using a Jain's fairness index-weighted
reward to address the fairness issue in LTE-unlicensed. In~\cite{xiao.zhan202208},
the author investigated the fairness of coexistence between unlicensed
nodes and Wi-Fi nodes by integrating 3GPP fairness. Tan \emph{et al}.~\cite{tan.niyato201905}
introduced the length of idle ending as an indicator of whether the
WiFi system has finished transmitting all buffered packets, incorporating
this indicator into the reward function to ensure fair coexistence
between license-assisted access LTE systems and WiFi systems. In~\cite{guo.sun202205},
the authors proposed a delay to last successful transmission (D2LT)
indicator and designed the reward function based on D2LT. The reward
function encourages the node with the largest delay to transmit, thereby
achieving proportional fairness among the nodes.  In~\cite{han.sun2024},
a reward function was defined to ensure fairness among nodes by assigning
additional rewards or penalties based on the sorted index of nodes'
throughput, prioritizing those with lower throughput. Additionally,
a new metric, the average age of a packet, was proposed in~\cite{jadoon.perez-cruz202204}
to measure the short-term imbalance among nodes, thereby ensuring
short-term fairness. However, most of these studies require prior
knowledge of the environments, such as the total number of  nodes
and the throughput of each node, which limits their applicability
in practical scenarios where such information is not available.

\section{System Model and Problem Formulation\label{sec:model}}

In this section, we first present the system model of the heterogeneous
wireless network and then formulate the multiple access problem as
an MDP.

\begin{figure}[!t]
\centering \includegraphics[width=0.8\columnwidth]{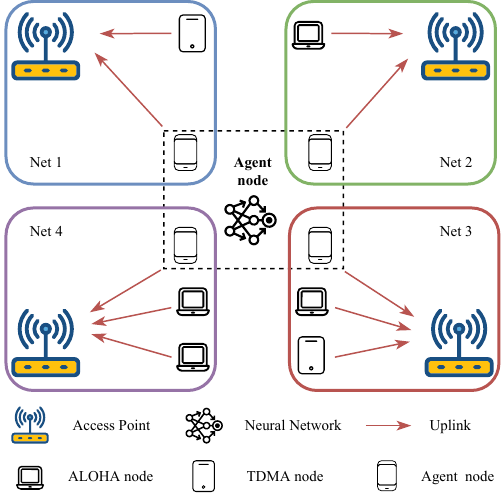}
\caption{Heterogeneous wireless networks with diverse coexisting scenarios.}
\label{fig:System-Model}
\end{figure}

\subsection{System Model}

We investigate a heterogeneous wireless network, where multiple nodes,
including $N$ existing nodes and an agent node, transmit packets
to an Access Point (AP) through a shared wireless channel. The set
of all nodes is denoted by $\{0,1,\ldots,N\}$, where $0$ represents
the agent node and $\{1,\ldots,N\}$ represents the existing nodes.
The network is considered heterogeneous because each node may utilize
a different MAC protocol. Importantly, there is no prior knowledge
available about the number of existing nodes or the MAC protocols
they employ. The system operates in a time-slotted manner, with each
time slot corresponding to the duration of a data packet transmission.
Nodes are  allowed to initiate transmission only at the beginning
of a time slot. A packet is considered successfully transmitted if
no other nodes transmit simultaneously, and in this case, the AP broadcasts
an ACK packet. However, if multiple nodes transmit concurrently in
the same time slot, a collision occur, and the AP broadcasts a NACK
packet to indicate an unsuccessful transmission.

In accordance with~\cite{yu.liew201906}, we consider four types
of time-slotted MAC protocols that may be used by  existing nodes:
$q$-ALOHA, Fixed-Window ALOHA (FW-ALOHA), Exponential Backoff ALOHA
(EB-ALOHA), and Time Division Multiple Access (TDMA). In $q$-ALOHA,
nodes transmit in each time slot with a fixed probability $q$. In
FW-ALOHA, a node generates a random counter in the range of $[0,W-1]$
after transmitting, and it must wait for the counter to expire before
initiating its next transmission. EB-ALOHA is a variant of FW-ALOHA
where the window size  doubles progressively when collisions occur
during transmission. This increase in window size continues until
a maximum size $2^{b}W$ is reached, where $b$ is the maximum backoff
stage. After  a successful transmission, the window size is reset
to its initial value $W$. TDMA divides time into frames, each consisting
of multiple time slots. Nodes are assigned specific time slots within
each frame according to a predetermined schedule.

The agent node decides whether to transmit data in each time slot.
Upon transmission, it receives immediate feedback from the AP, indicating
whether the transmission was successful or not. If the agent node
chooses not to transmit, it actively listens to the channel and gathers
observations about the environment. These observations provide information
about the transmission outcomes of other nodes or the idleness of
the channel. The agent node's objective is to maximize overall wireless
spectrum utilization across the entire network while ensuring harmonious
coexistence with existing nodes. This involves efficiently and fairly
utilizing the underutilized channels of existing nodes.

Considering the varying number of existing nodes and the potential
use of different MAC protocols, there is a wide range of coexisting
scenarios in heterogeneous wireless networks, as depicted in Fig.~\ref{fig:System-Model}.
We aim to design a generalizable MAC protocol that enables the coexistence
of the agent node with existing nodes across diverse heterogeneous
wireless networks. The protocol offers universal accessibility and
rapid adaptation capabilities, allowing the agent node to adjust to
different network conditions encountered in heterogeneous environments.

\subsection{MDP formulation}

We formulate the multiple access problem for the agent node in heterogeneous
wireless networks as an MDP. The components of the MDP are defined
as follows.

\subsubsection{Action}

The action of the agent node at the beginning of time slot $t$ is
denoted by $a_{t\text{,0}}\in\{0,1\}$. For simplicity, we will omit
the subscript 0 when it does not cause confusion. Here, $a_{t}=1$
represents the transmission of a packet at slot $t$, while $a_{t}=0$
indicates that the agent does not transmit at slot $t$ and instead
only listens to the channel.

\subsubsection{State}

After the agent takes action $a_{t}$, it receives a channel observation
$o_{t}\in\{0,1,2\}$ from the feedback broadcasted by the AP. Here,
\emph{$o_{t}=0$} indicates no transmission on the channel, \emph{$o_{t}=1$}
indicates that only one node transmitted on the channel, and \emph{$o_{t}=2$}
indicates a collision caused by simultaneous transmission from multiple
nodes. We denote the action-observation pair at time slot $t$ as
$h_{t}=\left(a_{t},o_{t}\right)$, with five possible combinations
for $h_{t}$, as summarized in Table~\ref{table:Obs}. Then, the
state of the agent node can be defined using the past action-observation
pairs in a history window to capture the temporal dynamics of the
environment. Specifically, the state at time slot $t$ is represented
as $s_{t}=\left[h_{t-L},h_{t-L+1},\ldots,h_{t-1}\right]$, where $L$
is the length of the state history window.

\begin{table}[!t]
\centering \caption{Possible action-observations pairs in each time slot. \label{table:Obs}}
\begin{tabular}{@{\hspace{0.2cm}}c@{\hspace{0.2cm}}c@{\hspace{0.2cm}}l}
\toprule
$a_{t}$ & $o_{t}$ & Description\tabularnewline
\midrule
0 & 0 & The channel is idle.\tabularnewline
0 & 1 & An existing node transmits successfully.\tabularnewline
0 & 2 & A collision occurs among existing nodes.\tabularnewline
1 & 1 & The agent node transmits successfully.\tabularnewline
1 & 2 & The agent node collides with existing nodes.\tabularnewline
\bottomrule
\end{tabular}
\end{table}

\subsubsection{Reward}

The primary goal of the agent node is to maximize the total throughput
of the entire network while ensuring fair coexistence with existing
nodes.  The  reward related to throughput at time slot $t$ is defined
as follows:
\begin{equation}
r_{t}^{p}=\begin{cases}
1, & \text{if }o_{t}=1,\\
0, & \text{otherwise.}
\end{cases}\label{eq:reward-max-throughput}
\end{equation}
This definition means that the agent node receives a reward when a
packet is successfully transmitted at time slot $t$, whether the
transmission is from the agent node or an existing node. However,
no reward is given in the case of a collision or when the channel
is idle.

To incorporate fairness considerations, we first define the short-term
average throughput of the agent node and the existing nodes. Let $S_{t,0}$
denote the short-term average throughput of the agent node  at time
slot $t$, which is defined as the ratio of successful transmissions
of the agent node within a window of $Z$ time slots, i.e.,  $S_{t,0}=\frac{1}{Z}\sum_{t^{\prime}=t-Z+1}^{t}r_{t^{\prime}}^{p}a_{t^{\prime}}$.
As the agent node cannot distinguish the actual short-term average
throughput of each existing node from the historical feedback provided
by the AP, we treat the existing nodes as a collective entity and
define the short-term average throughput of all existing node as $S_{t,N}=\frac{1}{Z}\sum_{t^{\prime}=t-Z+1}^{t}r_{t^{\prime}}^{p}(1-a_{t^{\prime}})$.
Then, we design a reward function that accounts for both throughput
and fairness  as follows:
\begin{equation}
r_{t}=r_{t}^{p}\cdot(1-\nu f_{t}),\label{eq:fair-reward}
\end{equation}
where
\begin{equation}
f_{t}=\begin{cases}
\frac{S_{t,0}}{S_{t,0}+S_{t,N}}, & \text{if }a_{t}=1,\\
\frac{S_{t,N}}{S_{t,0}+S_{t,N}}, & \text{otherwise},
\end{cases}\label{eq:fair-f}
\end{equation}
 and $\nu\in\left[0,1\right]$ is the fairness factor.

By adjusting the value of $\nu$, the reward function can balance
throughput maximization and fairness considerations. A larger value
of $\nu$ places greater emphasis on fairness, while a smaller value
prioritizes throughput. When $\nu=0$, the reward function focuses
solely on  the total network throughput. When $\nu=1$, the reward
can be rewritten as:
\begin{equation}
r_{t}=\begin{cases}
\frac{S_{t,N}}{S_{t,0}+S_{t,N}}, & \text{if }(a_{t},o_{t})=(1,1),\\
\frac{S_{t,0}}{S_{t,0}+S_{t,N}}, & \text{\text{if }\ensuremath{(a_{t},o_{t})=(0,1)}},\\
0, & \text{otherwise}.
\end{cases}\label{eq:fair-rf}
\end{equation}
When the agent node successfully transmits, the reward is proportional
to the ratio of the existing nodes' throughput to the total network
throughput. Conversely, if any existing node successfully transmits
while the agent node remains silent, the reward is proportional to
the ratio of the agent node's throughput to the total network throughput.
This reward mechanism promotes fairness by guiding the agent node
to seek transmission opportunities when its throughput is lower than
that of the existing nodes, and encouraging it to yield access when
its throughput exceeds that of existing nodes. In this way, the reward
function helps minimize the throughput differences between the agent
node and the existing nodes.

The  goal of a conventional reinforcement learning problem  is to
find an optimal policy $\pi^{*}$ that maximizes the expected sum
of discounted rewards for a given task, as expressed by:
\begin{equation}
\pi^{*}=\arg\max_{\pi}\mathbb{E}_{\text{\ensuremath{\tau}}}\left[\sum_{t}\gamma^{t}r\left(s_{t},a_{t}\right)\right],
\end{equation}
where $\gamma$ is the discount factor and $\tau$ is the  trajectory
induced by policy $\pi$. While the learned optimal policy performs
well for a specific training task, its  ability to generalize across
different tasks is limited. In meta-RL, the problem is extended to
a distribution of tasks $p(\mathcal{T})$. The training and test tasks
are both drawn from this distribution, where each task $\mathcal{T}^{k}$
corresponds to a different scenario, i.e., a distinct heterogeneous
wireless network configuration. The key idea in meta-RL is to learn
a policy that can generalize across a range of tasks, enabling the
agent to quickly adapt to new tasks without needing to start from
scratch.   The goal of  meta-RL is to find the optimal policy $\pi^{*}$
that maximize the expected cumulative discounted reward over   tasks
drawn from the task distribution, which is expressed as:
\begin{equation}
\pi^{*}=\arg\max_{\pi}\mathbb{E}_{\mathcal{T}^{k}\sim p(\mathcal{T})}\left[\mathbb{E}_{\text{\ensuremath{\tau^{k}}}}\left[\sum\limits _{t}\gamma^{t}r_{t}\right]\right],
\end{equation}
where $\tau^{k}$ is the  trajectory induced by policy $\pi$ for
task $\mathcal{T}^{k}$.

\section{Meta-RL-based GMA Protocol \label{sec:protocol}}

In this section, we first provide an overview of the proposed GMA
protocol. Next, we describe the neural networks utilized within GMA.
Finally, we outline the processes involved in both the meta-training
and meta-testing phases.
\begin{figure*}[!t]
\centering \includegraphics[width=0.95\textwidth]{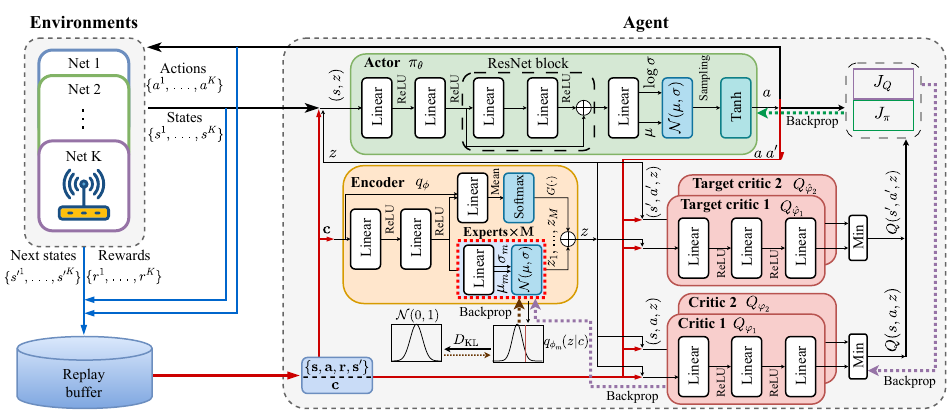}
\caption{The framework for GMA protocol.}
\label{fig:PEARL}
\end{figure*}

\subsection{Overview}

We propose the GMA protocol, which leverages meta-RL techniques with
MoE, to enhance the agent node\textquoteright s ability to rapidly
adapt and generalize across diverse heterogeneous network environments.
Building on the PEARL framework from~\cite{rakelly.quillen201905,retyk.retyk202104},
the GMA protocol includes an encoder~\cite{kingma.welling201405}
to generate task embeddings and uses SAC~\cite{haarnoja.levine201807,haarnoja.levine201901}
for learning a goal-conditioned policy. Unlike PEARL, we redesign
the encoder architecture by incorporating an MoE layer~\cite{Seniha.Paul2014},
which allows for the generation of mixture latent representations
from multiple experts. This enhancement enables the system to better
differentiate between various tasks.

 The encoder functions as an advanced feature extractor, distilling
task-specific information from a set of recent transitions (state,
action, reward, next state). Each transition is encoded using the
MoE layer, producing a set of rich and diverse representations. These
representations are then fused into a comprehensive task embedding,
which is provided to SAC as the conditioning input. The MoE-enhanced
encoder improves the system's ability to distinguish between different
tasks more effectively.  SAC operates both as a policy function and
an evaluation criterion. It utilizes an actor-critic architecture
with separate policy and value function networks to learn the meta-policy
using the task embeddings from the encoder. The actor generates task-specific
actions based on both the current state and task representations,
while the critic evaluates the performance of the conditioned policy.
This structure enables the agent to adapt more efficiently to varying
network conditions, thereby improving performance in meta-reinforcement
learning scenarios. Further details are provided in the following
subsections.

\subsection{Neural Networks}

 As depicted in Fig.~\ref{fig:PEARL}, the GMA protocol consists
of several key components: an MoE-enhanced encoder network  $q_{\phi}$,
an actor network $\pi_{\theta}$, two critic networks $Q_{\varphi_{1}}$
and $Q_{\varphi_{2}}$, and two target critic networks $Q_{\hat{\varphi}_{1}}$
and $Q_{\hat{\varphi}_{2}}$. Additionally, a replay buffer $B$ is
utilized to store the agent's transitions $(s,a,r,s^{\prime})$ and
facilitate the updating of network parameters.

\subsubsection*{Encoder network}

\begin{figure}[!t]
\centering \includegraphics[width=1\columnwidth]{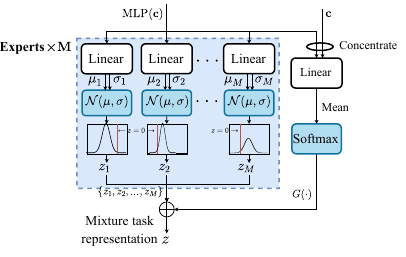}
\caption{MoE Probabilistic Embeddings.}
\label{fig:MoE-Encoder}
\end{figure}

The encoder network, parameterized by $\phi$, is responsible for
generating the task representation $z$ based on the recent context
$\bm{c}=\{c_{1},c_{2},...,c_{U}\}$, which comprises a set of transitions
$c_{u}=(s,a,r,s^{\prime})$ from the previous $U$ time steps. The
encoder combines a shared multilayer perceptron (MLP) backbone with
an MoE layer, which includes a gating layer $G$ and multiple experts
$m\in\{1,2,...,M\}$. The gating layer $G$ consists of a linear layer
followed by the Softmax activation to generates a set of weights:
\begin{equation}
G(\bm{c})=\text{Softmax}(\frac{1}{U}\sum_{u=1}^{U}\text{Linear}(\text{MLP}(c_{u}),c_{u})).\label{eq:softmax}
\end{equation}
As shown in \ref{fig:MoE-Encoder},  each expert $m$ is represented
by a linear layer $E_{m}$,  which independently encodes each individual
transition $c_{u}$ as a Gaussian factor $\varPsi_{\phi,{m}}(z_{m}\vert c_{u})=\mathcal{N}(\mu_{m}(c_{u}),\sigma_{m}(c_{u}))$.
Here, $\mu_{m}(c_{u})$ and $\sigma_{m}(c_{u})$ donate the mean and
variance of Gaussian factor, respectively. Given that the encoder
is permutation-invariant \cite{rakelly.quillen201905},  the Gaussian
factors derived from each transition can be multiplied together to
estimate the overall posterior of each expert $m$:
\begin{equation}
q_{\phi,m}(z\vert\bm{c})\propto\prod_{u=1}^{U}\varPsi_{\phi,m}(z\vert c_{u}).\label{eq:Gaussian-m}
\end{equation}
The task representation $z_{m}$ of each expert $m$ is then sampled
from $q_{\phi,m}(z\vert\bm{c})$. By weighting and combining the task
representations from all experts using the weights $G_{m}(\bm{c})$
generated by gating layer $G$, the final mixture task representation
is obtained as $z=\sum_{m=1}^{M}G_{m}(\bm{c})z_{m}$. The mixture
task representation $z$ serves as the conditioning input to SAC,
facilitating effective learning and adaptation. For simplicity, we
denote the overall process of task representation generation as $z\sim q_{\phi}(z\vert\bm{c})$.

\subsubsection*{Actor network}

The actor network, parameterized by $\theta$, generates actions $a$
based on the current state $s$ and the task representation $z$.
It consists of an input layer, a ResNet block with a shortcut connection~\cite{he2016},
and an output layer. The output of the actor network includes the
mean $\mu$ and the log-variance $\log\sigma$ for the Tanh-normal
distribution. From this distribution, we sample a continuous action
$\hat{a}$ that lies within the range of $-1$ to $1$. To obtain
a discrete action, we apply a simple binary mapping to  $\hat{a}$.
Specifically, the action $a$ is set to 1 if $\hat{a}\geq0$, and
0 otherwise.

\subsubsection*{Critic networks}

The two critic networks, parameterized by $\varphi_{1}$ and $\varphi_{2}$,
evaluate the value of an action taken in a given state and  task representation
by the actor. These critic networks are implemented as three-layer
MLPs, with each outputting a parameterized soft Q-function denoted
as:
\begin{equation}
Q_{\varphi_{i}}(s,a,z)=r+\gamma\mathbb{E}_{s^{\prime}\sim p(\cdot|s,a)}\left[V_{\varphi_{i}}(s^{\prime},z)\right],\label{eq:soft-q-1-1}
\end{equation}
 where  $p(\cdot|s,a)$ represents the state transition function.
The soft state value function is defined as:
\begin{equation}
V_{\varphi_{i}}(s,z)=\mathbb{E}_{a\sim\pi}\left[Q_{\varphi_{i}}(s,a,z)-\alpha\log\pi\left(a\vert s,z\right)\right]\label{eq:soft-value-1-1}
\end{equation}
where $\alpha$ is the temperature parameter that controls the trade-off
between policy entropy and reward. To mitigate overestimation of Q-values,
the soft Q-function is estimated by taking the minimum of the outputs
from the two critic networks, i.e., $\min\left(Q_{\varphi_{1}}(s,a,z),Q_{\varphi_{2}}(s,a,z)\right)$~\cite{hasselt.hasselt2010}.
Additionally, two target critic networks, parameterized by $\hat{\varphi}_{1}$
and $\hat{\varphi}_{2}$, are employed to reduce the correlation between
the target values and the current estimates of the critic networks,
thereby enhancing the stability of the training process. Similar to
the critic networks, the target critic networks are implemented as
three-layer MLPs.

\subsection{Procedure }

The procedure consists of two phases: meta-training and meta-testing.
In meta-training, the model learns a meta-policy by training on diverse
tasks. During meta-testing, the model quickly adapts to new, unseen
tasks using few-shot learning.

\subsubsection*{Meta-training}

\begin{algorithm}[t]
\begin{algorithmic}[1] \caption{Meta-Training of GMA Algorithm}
\label{alg:training}
\global\long\def\algorithmicrequire{\textbf{Require:}}%

\REQUIRE{Training task set $\{\mathcal{T}^{k}\}_{k=1,...,K}$ from
$p(\mathcal{T})$, learning rate $\lambda_{\phi},\lambda_{\pi},\lambda_{\alpha},\lambda_{Q}$,
soft update rate $\eta$, batch size $N_{E}$, collecting steps $N_{c}$}

\STATE Initialize the parameters $\phi,\theta,\varphi_{i},\hat{\varphi}_{i}\ (i=1,2)$
and $\alpha$

\STATE Initialize replay buffer $B^{k}$ for each task

\FOR {each episode}

\FORALL{$\mathcal{T}^{k}$}

\STATE Initialize environmental context $\bm{c}^{k}=\emptyset$

\FOR {$t=1,2,...,N_{c}$}

\STATE Sample $z^{k}\sim q_{\phi}(z^{k}\vert\bm{c}^{k})$

\STATE Gather transition $(s_{t},a_{t},r_{t},s_{t+1})$ from $\pi_{\theta}(a\vert s,z^{k})$
and store it into $B^{k}$

\STATE Sample $\bm{c}^{k}=\{(s_{u},a_{u},r_{u},s_{u}^{\prime})\}_{u=1,...,N_{E}}\sim B^{k}$

\ENDFOR

\ENDFOR

\FOR {each gradient step}

\STATE Sample context and transition batches from $\{B^{k}\}_{k=1,...,K}$

\STATE Sample $z^{k}\sim q_{\phi}(z^{k}\vert\bm{c}^{k})$ for all
$k=1,...,K$

\STATE Update networks:

\STATE $\varphi_{i}\leftarrow\varphi_{i}-\lambda_{Q}\nabla_{\varphi_{i}}\sum_{k}J_{Q}\left(\varphi_{i}\right)$,
$i=1,2.$

\STATE $\phi\leftarrow\phi-\lambda_{\phi}\nabla_{\phi}\sum_{k}J_{\text{en}}\left(\phi\right)$

\STATE $\theta\leftarrow\theta-\lambda_{\pi}\nabla_{\theta}\sum_{k}J_{\pi}\left(\theta\right)$

\STATE $\alpha\leftarrow\alpha-\lambda_{\text{temp}}\nabla_{\alpha}\sum_{k}J_{\text{temp}}\left(\alpha\right)$

\STATE $\hat{\varphi}_{i}\leftarrow\eta\varphi_{i}+(1-\eta)\hat{\varphi}_{i}$,
$i=1,2.$

\ENDFOR

\ENDFOR

\end{algorithmic}
\end{algorithm}

We consider a set of $K$ training tasks $\{\mathcal{T}^{k}\}_{k=1,...,K}$,
where each training task is drawn from the distribution $p(\mathcal{T})$.
In each training episode, there are two phases: the data collection
phase and the optimization phase. During the data collection phase,
the agent collects training data from different training tasks. Specifically,
for each training task, the agent leverages the encoder to derive
the task representation $z$ from the context $c$, which is sampled
uniformly from the most recently collected  data in the current episode.
Based on the task representation $z$ and the state $s$, the agent
generates an action $a$ using the actor network. The resulting transitions
are stored in individual replay buffers $B^{k}$ for each training
task $k$.

In the optimization phase, the loss functions for the encoder and
SAC are computed, and the parameters of the encoder, actor, and critic
networks are jointly updated using gradient descent. The critic networks
$Q_{\varphi_{1}}$ and $Q_{\varphi_{2}}$ are updated by minimizing
the soft Bellman error:
\begin{equation}
\begin{aligned}J_{Q}\left(\varphi_{i}\right)= & \underset{\;z\sim q_{\phi}(z|\bm{c})}{\mathbb{E}_{(s,a,r,s^{\prime})\sim B}}\bigg[Q_{\varphi_{i}}(s,a,z)-\\
 & \big(r+\gamma(\min\limits _{i=1,2}Q_{\hat{\varphi}_{i}}\left(s^{\prime},a^{\prime},z\right)-\alpha\log\pi_{\theta}\left(a^{\prime}\vert s^{\prime},z\right))\big)\bigg]^{2}.
\end{aligned}
\label{eq:Critic-loss}
\end{equation}
The target critic networks $Q_{\hat{\varphi}_{1}}$ and $Q_{\hat{\varphi}_{2}}$are
updated by:
\begin{equation}
\hat{\varphi}_{i}\leftarrow\eta\varphi_{i}+(1-\eta)\hat{\varphi}_{i},\quad i=1,2,\label{eq:soft-update}
\end{equation}
where $\eta$ is the soft update rate.

The encoder's loss is defined based on the variational lower bound,
which consists of two key components: the soft Bellman error for the
critic and the sum of the Kullback-Leibler (KL) divergence for each
expert. Specifically, the loss function for the encoder is given by:
\begin{equation}
\begin{aligned}J_{\text{en}}\left(\phi\right)=\mathbb{E}_{\mathcal{T}}\bigg[\mathbb{E}_{z\sim q_{\phi}(z|\bm{c})}\bigg[ & \sum_{i=1}^{2}J_{Q}\left(\varphi_{i}\right)+\\
 & \beta\sum_{m=1}^{M}D_{\text{KL}}\bigl(q_{\phi,m}(z\vert\bm{c})\Vert p(z)\bigr)\bigg]\bigg],
\end{aligned}
\label{eq:encoder-loss}
\end{equation}
where $p(z)$ is a unit Gaussian prior, $D_{\text{KL}}(\cdot||\cdot)$
is the KL divergence, and $\beta$ is the weight the KL-divergence
term. The KL divergence term for each expert is used to constrain
the mutual information between the task representation $z$ and the
context $c$, ensuring that $z$ contains only relevant information
from the context and mitigating overfitting.

The  loss utilized to update the actor network parameters is given
by:
\begin{equation}
J_{\pi}\left(\theta\right)=\underset{\;\;z\sim q_{\phi}(z|\bm{c})}{\mathbb{E}_{s\sim B,a\sim\pi_{\theta}}}\left[\alpha\log(\pi_{\theta}\left(a\vert s,z\right))-\min\limits _{i=1,2}Q_{\varphi_{i}}(s,a,z)\right].\label{eq:Actor-loss}
\end{equation}
The temperature parameter $\alpha$ is updated using the following
loss function:
\begin{equation}
J_{\text{temp}}\left(\alpha\right)=\underset{\;z\sim q_{\phi}(z|\bm{c})}{\mathbb{E}_{s\sim B,a\sim\pi_{\theta}}}\left[-\alpha\log(\pi_{\theta}\left(a\vert s,z\right))-\alpha\bar{\mathcal{H}}\right],\label{eq:Alpha-loss}
\end{equation}
where $\bar{\mathcal{H}}$ represents a target expected entropy, which
is set to $-\dim(\mathcal{A})$~\cite{haarnoja.levine201901}, and
$\mathcal{A}$ denotes the action space. The details of the meta-training
process are summarized in Algorithm~\ref{alg:training}.

\subsubsection*{Meta-testing}

\begin{algorithm}[t]
\begin{algorithmic}[1] \caption{Meta-Testing With Fine-Tuning }
\label{alg:fine-tune}
\global\long\def\algorithmicrequire{\textbf{Require:}}%

\REQUIRE{Testing task $\mathcal{T}^{k^{\prime}}$ from $p(\mathcal{T})$,
fine-tuning time step set $T_{\text{ft}}$, meta-trained parameters
$\theta$, $\phi$, $\varphi_{i}$, $\hat{\varphi}_{i}\ (i=1,2)$,
learning rate $\lambda_{\pi},\lambda_{\alpha},\lambda_{Q}$, soft
update factor $\eta$, batch size $N_{E}$, collecting steps $N_{c}$,
context buffer size $U$}

\STATE Initialize replay buffers $B^{k^{\prime}}$, context $\bm{c}^{k^{\prime}}=\emptyset$,
$\alpha$, $\theta^{k^{\prime}}=\theta$, $\varphi_{i}^{k^{\prime}}=\varphi_{i}$,
$\hat{\varphi}_{i}^{k^{\prime}}=\hat{\varphi}_{i}\ (i=1,2)$

\FOR {$t=1,2,...,3N_{c}$}

\STATE Sample $z^{k^{\prime}}\sim q_{\phi}(z^{k^{\prime}}\vert\bm{c}^{k^{\prime}})$

\STATE Gather transition $(s_{t},a_{t},r_{t},s_{t+1})$ from $\pi_{\theta^{k^{\prime}}}(a\vert s,z^{k^{\prime}})$
and store it into $B_{k}^{\prime}$

\STATE Update context $\bm{c}^{k^{\prime}}$ with transition $(s_{t},a_{t},r_{t},s_{t+1})$

\ENDFOR

\FOR {remain episode length}

\STATE Sample $z^{k^{\prime}}\sim q_{\phi}(z^{k^{\prime}}\vert\bm{c}^{k^{\prime}})$

\STATE $a_{t}\sim\pi_{\theta^{k^{\prime}}}(a_{t}\vert s_{t},z^{k^{\prime}})$

\STATE Execute $a_{t}$, receive $s_{t+1}\sim p(\cdot|s_{t},a_{t})$
and $r_{t}$

\STATE Update context $\bm{c}^{k^{\prime}}$ with transition $(s_{t},a_{t},r_{t},s_{t+1})$

\STATE Store transition $(s_{t},a_{t},r_{t},s_{t+1})$ into $B^{k^{\prime}}$

\IF {$t\text{\ensuremath{\in T_{\text{ft}}}}$}

\STATE Sample transition batch from $B^{k^{\prime}}$

\STATE Update networks:

\STATE $\varphi_{i}^{k^{\prime}}\leftarrow\varphi_{i}^{k^{\prime}}-\lambda_{Q}\nabla_{\varphi_{i}^{k^{\prime}}}J_{Q}\left(\varphi_{i}^{k^{\prime}}\right)$,
$i=1,2.$

\STATE $\theta^{k^{\prime}}\leftarrow\theta^{k^{\prime}}-\lambda_{\pi}\nabla_{\theta^{k^{\prime}}}J_{\pi}\left(\theta^{k^{\prime}}\right)$

\STATE $\alpha\leftarrow\alpha-\lambda_{\text{temp}}\nabla_{\alpha}J_{\text{temp}}\left(\alpha\right)$

\STATE $\hat{\varphi}_{i}^{k^{\prime}}\leftarrow\eta\varphi_{i}^{k^{\prime}}+(1-\eta)\hat{\varphi}_{i}^{k^{\prime}}$,
$i=1,2.$

\ENDIF

\ENDFOR

\end{algorithmic}
\end{algorithm}

In the meta-testing phase, the agent fine-tunes the actor and critic
networks to adapt the policy to new tasks in a few-shot manner. The
parameters of the encoder are fixed during this phase. When encountering
a new task, the agent collects context data over a few time steps.
The first transition is collected with task representation $z$ sampled
from the prior $q_{\phi}(z)=p(z)$.   Subsequent transitions are collected
with $z\sim q_{\phi}(z\vert\bm{c})$, where the context $\bm{c}$
is updated by selecting the latest $U$ transitions. As more context
is accumulated, the task representation $z$ becomes increasingly
accurate. The actor and critic networks are then updated using a procedure
similar to that in the meta-training phase, enabling the agent to
rapidly improve its performance on the new task. The details of the
meta-testing process are summarized in Algorithm~\ref{alg:fine-tune}.

\section{Performance Evaluation\label{sec:results}}

In this section, we conduct extensive simulations using PyTorch to
thoroughly evaluate the performance of the proposed GMA protocol.
We first describe the simulation setup and then compare the performance
of the proposed GMA protocol with two baseline protocols in different
scenarios.

\subsection{Simulation Setup}

\subsubsection{Hyperparameters}

The  state history length $L$ is set to 20, the  short-term average
throughput window $Z$ is set to 500, and the discount factor $\gamma$
is set to 0.9. Unless otherwise specified, the fairness factor $\nu$
is set to a default value of 0. For the encoder network, the   KL
divergence weight $\beta$ is set to 1, the default number of experts
$M$ is set to 3, and the  task representation dimension is set to
6. A replay buffer of  size  1000 is used to store past experiences.
 Both transition and context batches are sampled with a batch size
$N_{E}$ of 64, and the context buffer size $U$ is set to 150. Each
 hidden layer in the encoder, actor, and critic networks consists
of 64 neurons with  ReLU activation. The Adam optimizer with a learning
rate of 0.003 is used to optimize all trainable parameters, and the
soft update rate $\eta$ is set to 0.005. During the meta-training
phase, we collect data for a total of 200 steps for the GMA protocol
($N_{c}=200$), while during the testing phase, we collect data for
50 steps ($N_{c}=50$).

\subsubsection{Performance Metrics}

To evaluate the performance, we adopt  $S_{t}=S_{t,0}+S_{t,N}$ as
the  throughput metric.  We also employ  Jain's index~\cite{jain1984quantitative}
as a fairness metric in Section \ref{subsec:fair-coexistence}. Jain\textquoteright s
index is defined as $\frac{S_{t,0}^{2}+S_{t,N}^{2}}{2(S_{t,0}^{2}+S_{t,N}^{2})}$,
which quantifies the fairness of throughput distribution between the
agent node and existing nodes. For consistency, all simulation results
are averaged over 10 independent experiments per scenario, and Jain's
index is computed based on these averaged short-term throughput values.

\subsubsection{Baseline Protocols}

We compare the performance of the proposed GMA protocol with the following
baseline protocols:
\begin{itemize}
\item DLMA: The agent node accesses the channel using the DLMA protocol~\cite{yu.liew201906}.
\item DLMA-SAC: This baseline is a variant of DLMA, where the original DQN
component is replaced with the SAC.
\item Vanilla GMA: A specific variant of GMA, where the number of experts
$M$ is set to 1~\cite{liu.chen2023}.
\item Optimal policy: This baseline assumes that the agent node has complete
knowledge of the MAC mechanisms employed by existing nodes in the
network, providing the optimal network throughput as the benchmark
\cite{yu.liew201906}.
\end{itemize}

To ensure a fair comparison, all algorithms are designed with neural
networks of similar parameter counts for decision making. This allows
us to focus on the differences in their learning algorithms and evaluate
their relative performance in a consistent manner.

\subsubsection{Notations}

To distinguish between different scenarios in heterogeneous wireless
networks, we introduce a notation system  based on the protocols and
settings employed. The following notations are defined:
\begin{itemize}
\item $q$-ALOHA($q$): A scenario with a single existing node using the
$q$-ALOHA protocol for access, where the transmission probability
 is denoted by $q$.
\item FW-ALOHA($W$): A scenario with a single existing node using the FW-ALOHA
protocol, where the window size is fixed at $W$.
\item EB-ALOHA($W$): A scenario where a single existing node employs  the
EB-ALOHA protocol, with an initial window size $W$ and a maximum
backoff stage $b=2$.
\item TDMA($X$): A scenario where a single existing node transmits  in
 the specified slot $X$ within a frame consisting of 10 slots.
\end{itemize}

\subsection{Meta-training Performance}

\label{subsec:meta-training}
\begin{figure}[!t]
\centering \includegraphics[width=3in]{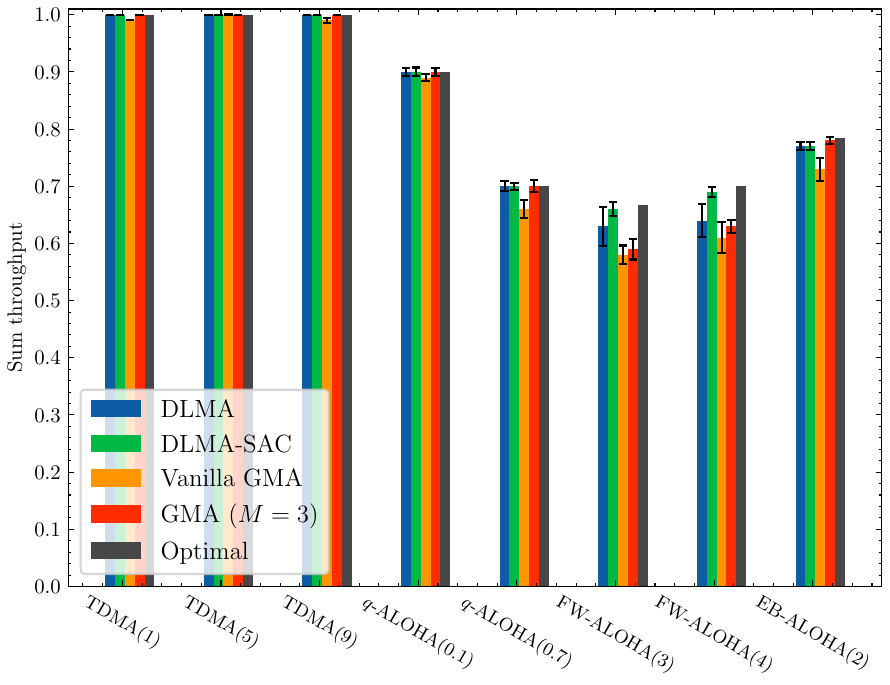}\caption{Comparison of the performance of different protocols in various training
environments. Each experiment was conducted over 20,000 time slots,
with error bars representing the standard deviations from 10 simulations
per case. }
\label{fig:Meta-Trained-result}
\end{figure}

\begin{figure*}[!t]
\centering \includegraphics[width=1\textwidth]{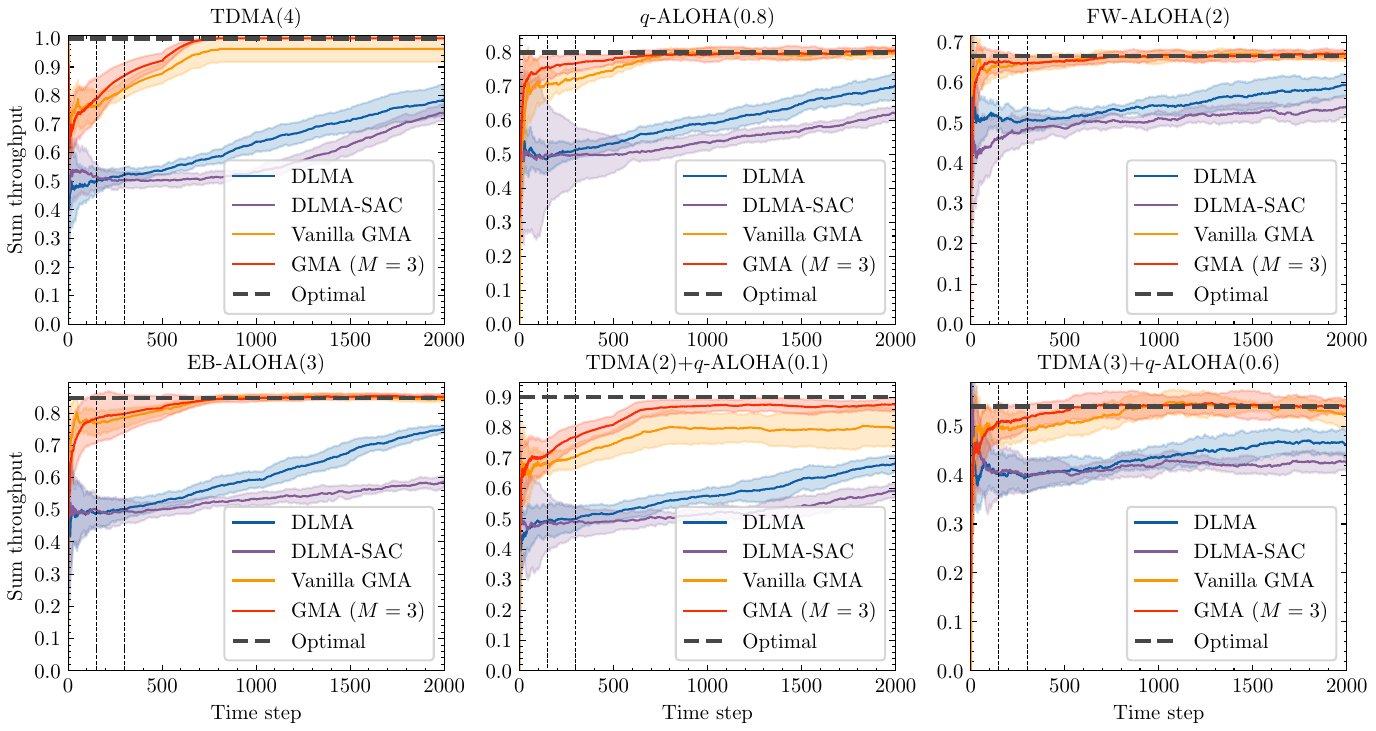}\caption{Comparison of the performance of different protocols in various testing
environments. The first vertical dashed line at the 150th time step
marks the beginning of agents' fine-tuning or training, while the
second vertical dashed line indicates the completion of the fine-tuning
 for the GMA agent. The DLMA and DLMA-SAC agents continue updating
beyond this point. The shaded area around the curves represents the
standard deviations from 10 experiments.}
\label{fig:Meta-test-result}
\end{figure*}

A set of training tasks is employed for meta-training. Specifically,
eight distinct environments are considered, including three TDMA environments
with time slots $X$ of 1, 5, and 9, two $q$-ALOHA environments with
transmission probabilities $q$ of 0.1 and 0.7, two FW-ALOHA environments
with window sizes $W$ of 3 and 4, and one EB-ALOHA environment with
an initial window size $W$ of 2.

The performance of different MAC protocols in various training environments
is shown in Fig.~\ref{fig:Meta-Trained-result}. The DLMA and DLMA-SAC
nodes are trained from scratch in each environment during the training
phase, while both the vanilla GMA and GMA node are trained on all
environments in the training task set. For simplicity, we refer to
both versions of GMA (vanilla and with multiple experts) as \textquotedbl GMA\textquotedbl{}
in subsequent references. The simulation results demonstrate that
the GMA protocol performs well across multiple environments,  with
only slight performance degradation compared to the DLMA and DLMA-SAC
protocols, which are optimized for specific environments. In addition,
GMA with additional experts provides more stability and higher performance
than the vanilla GMA due to its improved representation.

\subsection{Meta-testing Performance}

\label{subsec:meta-test}

In the testing phase, we evaluate the generalization capabilities
of the proposed GMA protocol on new tasks that were not encountered
during the training phase. We consider a total of six environments:
 four  with a single  coexisting nodes and two  with multiple  coexisting
nodes. The single-node environments are TDMA(5), $q$-ALOHA(0.8),
FW-ALOHA(2), and EB-ALOHA(3). These environments represent different
MAC protocols and settings, including TDMA with a time slot of 5,
$q$-ALOHA with a transmission probability of 0.8, FW-ALOHA with a
window size of 2, and EB-ALOHA with a window size of 3. The multi-node
environments include one  with TDMA(2) and $q$-ALOHA(0.1), and another
 with TDMA(3) and $q$-ALOHA(0.6). These environments involve  combinations
of TDMA and $q$-ALOHA protocols with different time slots and transmission
probabilities.

In each testing environment,  agents collect transitions during the
first 150 time steps to fill the replay buffer. During this period,
the GMA agent also accumulates context information. After  150 time
steps, all agents begin the fine-tuning or training phase. The GMA
agent is updated every 50 time steps, while the DLMA and DLMA-SAC
agents are updated every 5 time steps. After the 300th time step,
the GMA agent completes its fine-tuning, while the DLMA and DLMA-SAC
agents continue  updating. Note that all hyperparameters used during
the testing phase are consistent with those used in the training phase.

As illustrated in Fig.~\ref{fig:Meta-test-result}, the GMA agent
exhibits rapid convergence towards a near-optimal access strategy
after just three updates in all testing environments. In contrast,
the DLMA and DLMA-SAC agents  perform poorly at first and fail to
converge within the given time duration, despite undergoing ten times
as many updates. This highlights the GMA protocol's significant advantage
in adapting to new environments compared to previous DRL protocols
trained from scratch. Moreover, the GMA agent achieves higher sum
throughput compared to the other two baseline protocols. This is because
the GMA agent leverages the task representation extracted by the encoder
and utilizes the knowledge learned from previous environments to quickly
identify the optimal access strategy. Although the GMA agent was trained
on simpler tasks with single-node environments, it shows remarkable
initial performance and few-shot capabilities when handling multi-node
environments.  Additionally, due to the incorporation of the MoE architecture,
the GMA achieves higher initial performance and converges faster compared
to the vanilla GMA. It indicates that the MoE architecture enhances
representation learning, allowing the agent to better leverage prior
knowledge and adapt more efficiently to new environments.

\begin{figure}[t]
\centering \includegraphics[width=0.8\columnwidth]{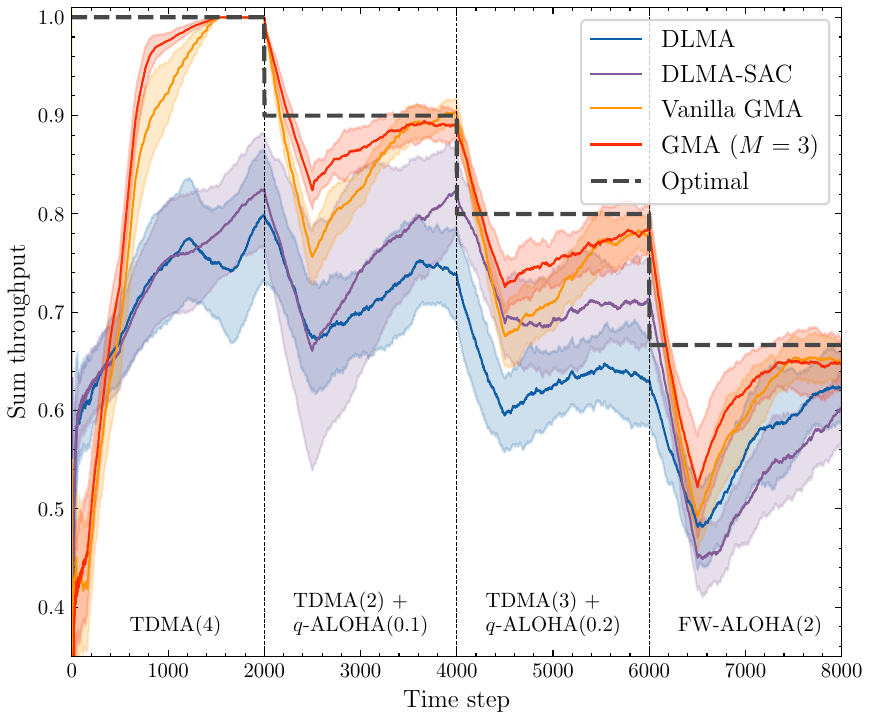}
\caption{Sum throughput in a dynamic environment. }
\label{fig:continue-result-set2}
\end{figure}

Next, we evaluate the rapid adaptability of the proposed GMA protocol
in a dynamic environment, where the number of existing nodes and their
protocols change every 2000 time slots. Both DLMA and DLMA-SAC agents
are trained in the TDMA(7) environment, while the GMA agent is initialized
with the  model obtained from meta-training. All agents are updated
every 50 time steps. After each environmental change, GMA  updates
only 16 times, whereas DLMA and DLMA-SAC  continue updating. The initial
scenario we investigate involves the agent coexisting  with one TDMA(4)
node. At the 2000th time slot, a $q$-ALOHA node with a transmission
probability $q$ of 0.1 joins, the agent must coexist with one $q$-ALOHA
node  and one TDMA node with a time slot $2$. At the 4000th time
slot, the  assigned slot $X$ for TDMA changes from 2 to 3, and the
transmission probability of $q$-ALOHA increases from 0.1 to 0.2.
At the 6000th time slot, the TDMA and $q$-ALOHA nodes leave, and
one FW-ALOHA node with a window size of 2 joins. As depicted in Fig.~\ref{fig:continue-result-set2},
our protocol quickly adapts to environmental changes and re-learns
a near optimal access strategy. While the DLMA and DLMA-SAC fail to
track the environmental changes and need more updates to converge,
it demonstrates that our protocol achieves significantly faster convergence
in a dynamic environment and can rapidly adapt to changes. Furthermore,
compared to the vanilla GMA, the GMA with three experts  adapts more
rapidly, showing the benefits of expert diversity.

\subsection{The Impact of the Size of the Meta-training Set}

\begin{figure*}[!t]
\centering \includegraphics[width=0.98\textwidth]{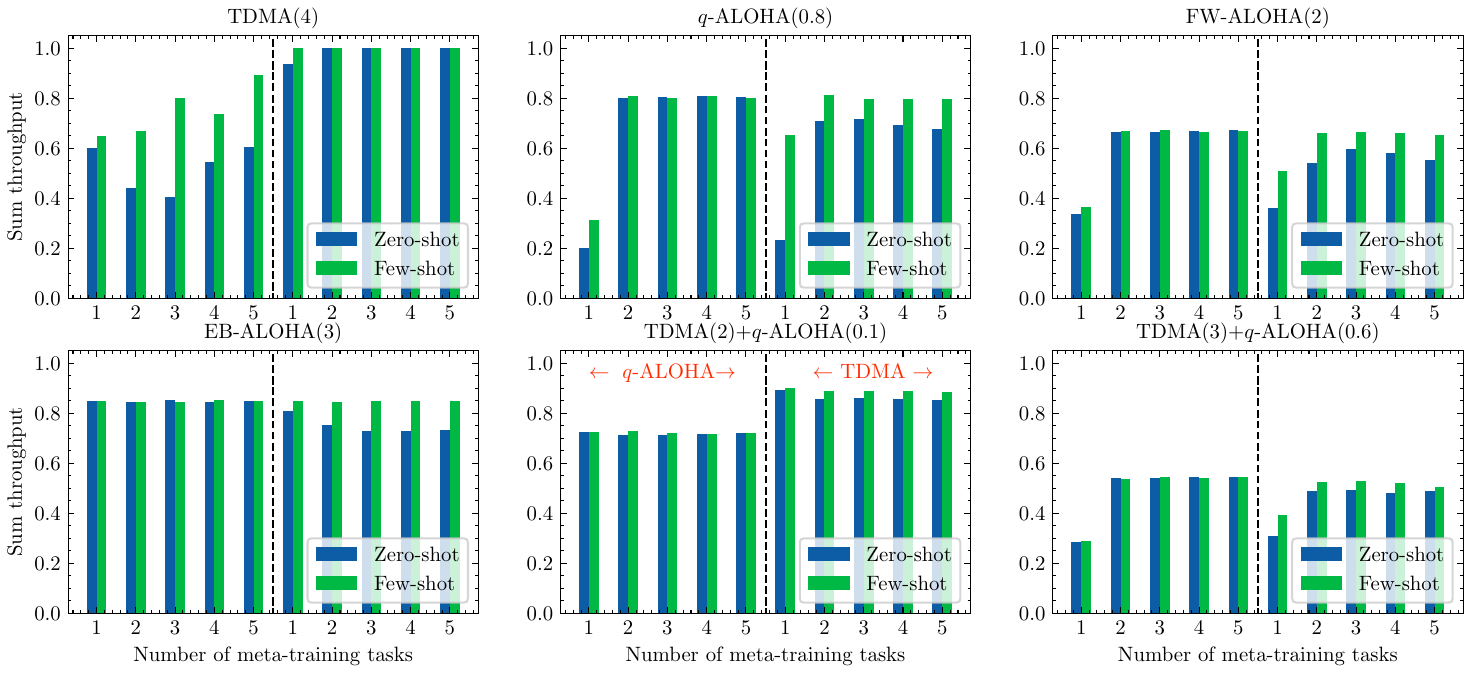}
\caption{A performance comparison with an increasing  number of meta-training
tasks. The left side of the vertical dashed line represents a set
of  $q$-ALOHA tasks, while the right side represents a set of  TDMA
tasks. Each meta-training  set consists of distinct tasks, and the
$x$-axis represents the size of the  set.}
\label{fig:tasks-num}
\end{figure*}

In this subsection, we analyze the generalization capabilities of
the proposed GMA protocol by varying the size of the meta-training
sets.  Specifically, as illustrated in Fig.~\ref{fig:tasks-num},
we investigate the impact of increasing the number of $q$-ALOHA tasks
and TDMA tasks within these sets. On the left side of the figures
(before the vertical dashed line), we present a series of $q$-ALOHA
environments with transmission probabilities $q$ of 0.1, 0.7, 0.5,
0.3, and 0.9. The $x$-axis indicates the cumulative inclusion of
environments in the meta-training set (e.g., the value of 2 on the
$x$-axis corresponds to the first two environments with $q$ values
of 0.1 and 0.7). Correspondingly, on the right side of the vertical
dashed line, we investigate multiple TDMA environments with time slots
$X$ of 1, 9, 5, 3, and 7, following the same $x$-axis representation.
As in Section~\ref{subsec:meta-test}, we assess the influence of
the training set size  by evaluating adaptation performance across
six distinct environments during the meta-testing phase.

Fig.~\ref{fig:tasks-num} illustrates the zero-shot and few-shot
performance  as the number of training tasks increases. In general,
an increase in the number of tasks during meta-training leads to an
overall improvement in zero-shot performance. However, there is a
threshold beyond which further gains are limited due to model size
constraints. As a result, the performance might fall short of the
level achieved by the optimal strategy. Comparing the adaptation performance
in the six distinct environments between the TDMA and $q$-ALOHA meta-training
sets, we observe that the composition of the meta-training set directly
influences zero-shot performance. The meta-trained model shows higher
generalization performance when the adaptation tasks closely resemble
those in the training set. Furthermore, Fig.~\ref{fig:tasks-num}
shows that for environments with TDMA nodes,  models meta-trained
on  pure TDMA sets outperform those trained on  $q$-ALOHA sets. Conversely,
for environments with only ALOHA nodes,  models meta-trained on  pure
$q$-ALOHA sets perform better than those trained on TDMA sets. For
environments with multiple existing nodes, the environment features
tend to be dominated by the transmission characteristics of a certain
dominant node. Therefore, the similarity between the meta training
task set and this dominant node determines the testing performance
in such environments. Additionally, with three-step fine-tuning,  models
meta-trained on pure TDMA sets achieve near-optimal performance across
various environments, while models meta-trained on pure $q$-ALOHA
sets may fail to converge in some environments. The GMA protocol demonstrates
the ability to rapidly generalize to new tasks, even with a smaller
number of tasks in the training set, reducing sampling and model update
costs.

\subsection{The Impact of the Diversity of the Meta-training Set}

\begin{figure*}[!t]
\centering \includegraphics[width=0.98\textwidth]{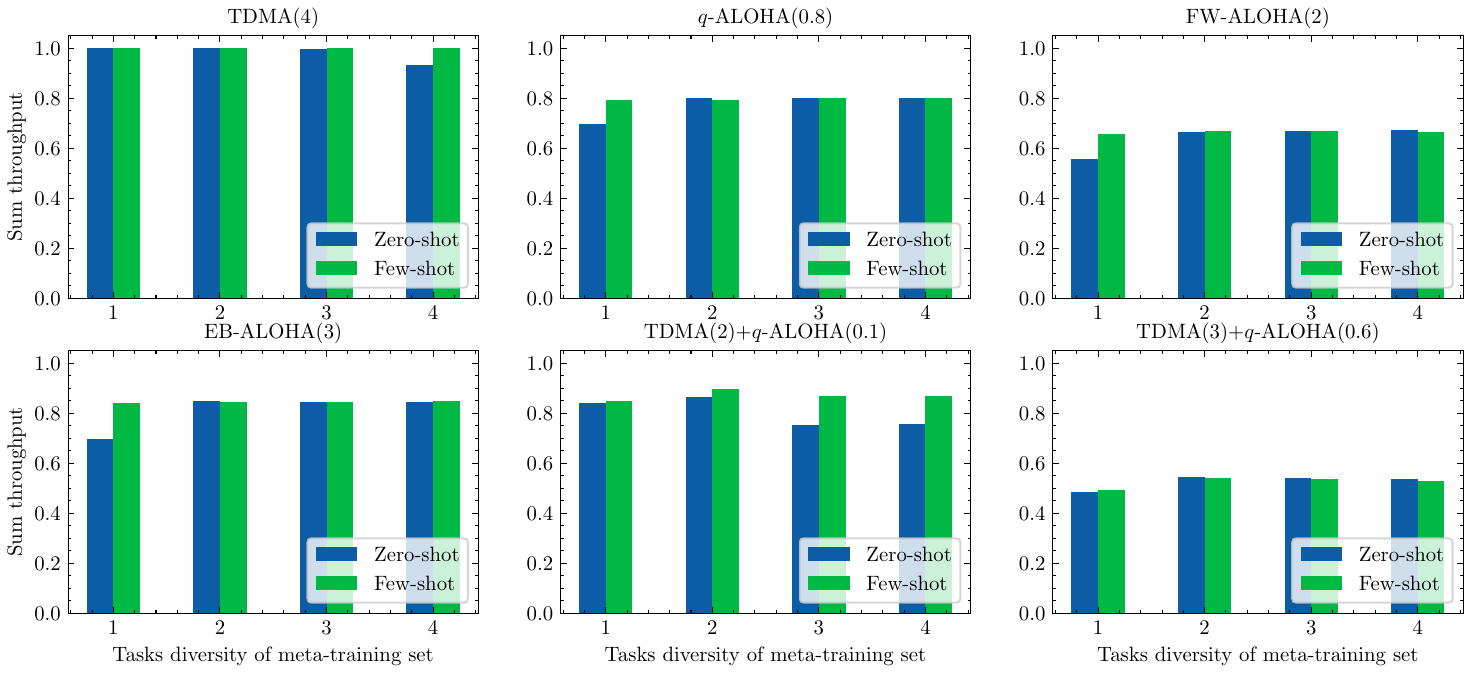}
\caption{Meta-training with gradual increases in the diversity of training
tasks set, the $x$-axis represents the set index.}
\label{fig:tasks-diversity}
\end{figure*}

In this subsection, we evaluate the generalization capabilities of
the proposed GMA protocol in terms of performance, considering the
diversity of tasks in the meta-training sets. The meta-training sets
consist of a fixed number of 8 distinct tasks. The sets  in Fig.~\ref{fig:tasks-diversity}
are defined as follows:
\begin{itemize}
\item Set 1: Eight TDMA environments with time slots $X$ of 1, 2, 3, 5,
6, 7, 8, and 9 are considered.
\item Set 2: Six TDMA environments with time slots $X$ of 1, 3, 5, 6, 7,
and 9, and two $q$-ALOHA environments with transmission probabilities
$q$ of 0.1 and 0.7 are considered.
\item Set 3: Five TDMA environments with time slots $X$ of 1, 5, 6, 7,
and 9, two $q$-ALOHA environments with transmission probabilities
$q$ of 0.1 and 0.7, and one EB-ALOHA environment with a window size
$W$ of 2 are considered.
\item Set 4: Three TDMA environments with time slots $X$ of 1, 5, and 9,
two $q$-ALOHA environments with transmission probabilities $q$ of
0.1 and 0.7, two FW-ALOHA environments with window sizes $W$ of 3
and 4, and one EB-ALOHA environment with a window size $W$ of 2 are
considered.
\end{itemize}

As the set index increases, the diversity of the meta-training sets
also increases. By examining the adaptation performance of the GMA
protocol across these sets, we can gain insights into how the diversity
and complexity of tasks impact the generalization abilities of the
model. As in Section~\ref{subsec:meta-test}, we evaluate the impact
of task diversity by analyzing the adaptation performance across six
distinct environments during the meta-testing phase.

As shown in Fig.~\ref{fig:tasks-diversity}, increasing the diversity
of the meta-training set during the meta-training stage leads to improved
initial performance in a larger number of environments during the
meta-testing stage. This can be attributed to the fact that a more
diverse meta-training set allows the encoder to learn to differentiate
between environments with different types of features and develop
a more comprehensive understanding of how to interact with diverse
features. Consequently, higher diversity facilitates better matching
of features in new environments during the meta-testing stage, enabling
the agent to leverage similar feature experiences for improved adaptation.
However, it is important to note that increased diversity does not
always benefit specific environments. For instance, in the case of
TDMA(4), the zero-shot performance actually decreases as the diversity
increases. This is because although the overall diversity increases,
the diversity specifically within the \textquotedbl TDMA\textquotedbl{}
category decreases. This aligns with the expected results discussed
earlier. Therefore, the measure of diversity in the meta-training
set should be carefully considered based on the distribution of different
environments that need to be adapted to. Additionally, the selection
of the meta-training set directly impacts the cost of deployment.
Increasing diversity typically comes at the expense of increased sampling
and model update costs. Therefore, it's crucial to strike a balance
between diversity and practical considerations such as resource constraints
and deployment efficiency.

\subsection{The Impact of the Number of Experts\label{subsec:expert-ablation}}

\begin{figure}[tb]
\centering \includegraphics[width=0.8\columnwidth]{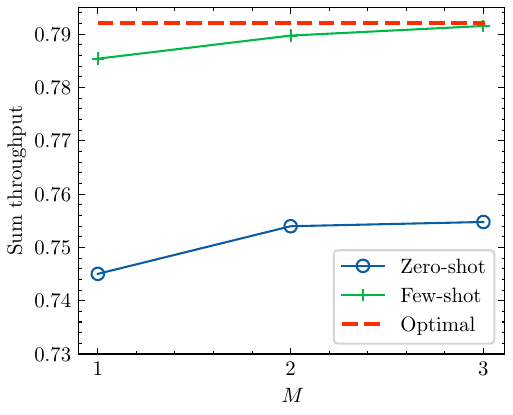}
\caption{GMA with different number of experts $M$.}
\label{fig:experts}
\end{figure}

\begin{figure}[h]
\centering \includegraphics[width=0.75\columnwidth]{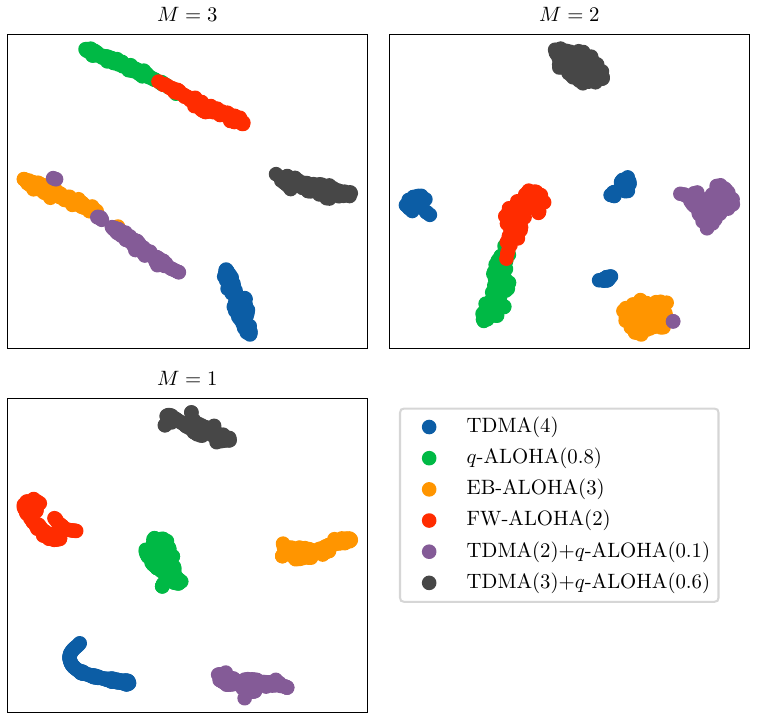}
\caption{The t-SNE visualization of latent representations extracted from trajectories
collected in the meta-testing environments with different number of
experts $M$.}
\label{fig:experts-tsne}
\end{figure}

In this subsection, we investigate the effect of the number of experts,
$M$. Both the training set and testing set are the same as those
described in Sections~\ref{subsec:meta-training} and \ref{subsec:meta-test}.
As shown in Fig.~\ref{fig:experts}, both zero-shot and few-shot
performance improve as the number of experts $M$ increases, reaching
near-optimal performance when $M=3$.

To better understand this effect, we analyze the behavior of the agent
from the perspective of dominant nodes. In environments such as $q$-ALOHA(0.8),
FW-ALOHA(2), and \textquotedbl TDMA(3)+$q$-ALOHA(0.6)\textquotedbl ,
the existing nodes dominate the overall throughput \cite{yu.liew201906},
which results in the agent node opting not to transmit, as it aims
to avoid contention. Conversely, in \textquotedbl TDMA(2)+$q$-ALOHA(0.1)\textquotedbl{}
and EB-ALOHA(3), the agent node is responsible for the majority of
the throughput and, as a result, transmits more frequently to maximize
performance \cite{yu.liew201906}. As illustrated in Fig.~\ref{fig:experts-tsne},
when $M=1$, similar behavior patterns of the agent in different environments
are only loosely grouped, indicating a lack of precision in the learned
representation. However, as $M$ increases, the representation becomes
more distinct and well-clustered, suggesting that the use of additional
experts allows for better differentiation between various behaviors.
This improvement in clustering reflects enhanced representation learning,
which directly contributes to the agent's ability to adapt effectively
across diverse network environments.

\subsection{Fair Coexistence Objective with Existing Nodes\label{subsec:fair-coexistence}}

\begin{figure}[t]
\centering \includegraphics[width=0.9\columnwidth]{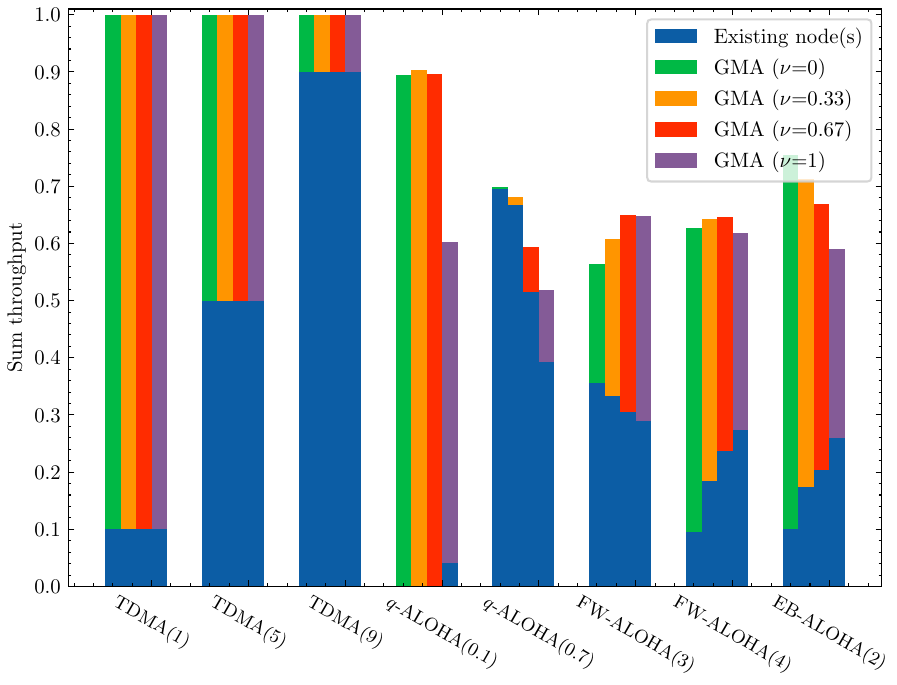}

\caption{Meta-training with different fairness factor $\nu$.}
\label{fig:fairness-train}
\end{figure}

\begin{figure*}[tbh]
\centering \includegraphics[width=0.95\textwidth]{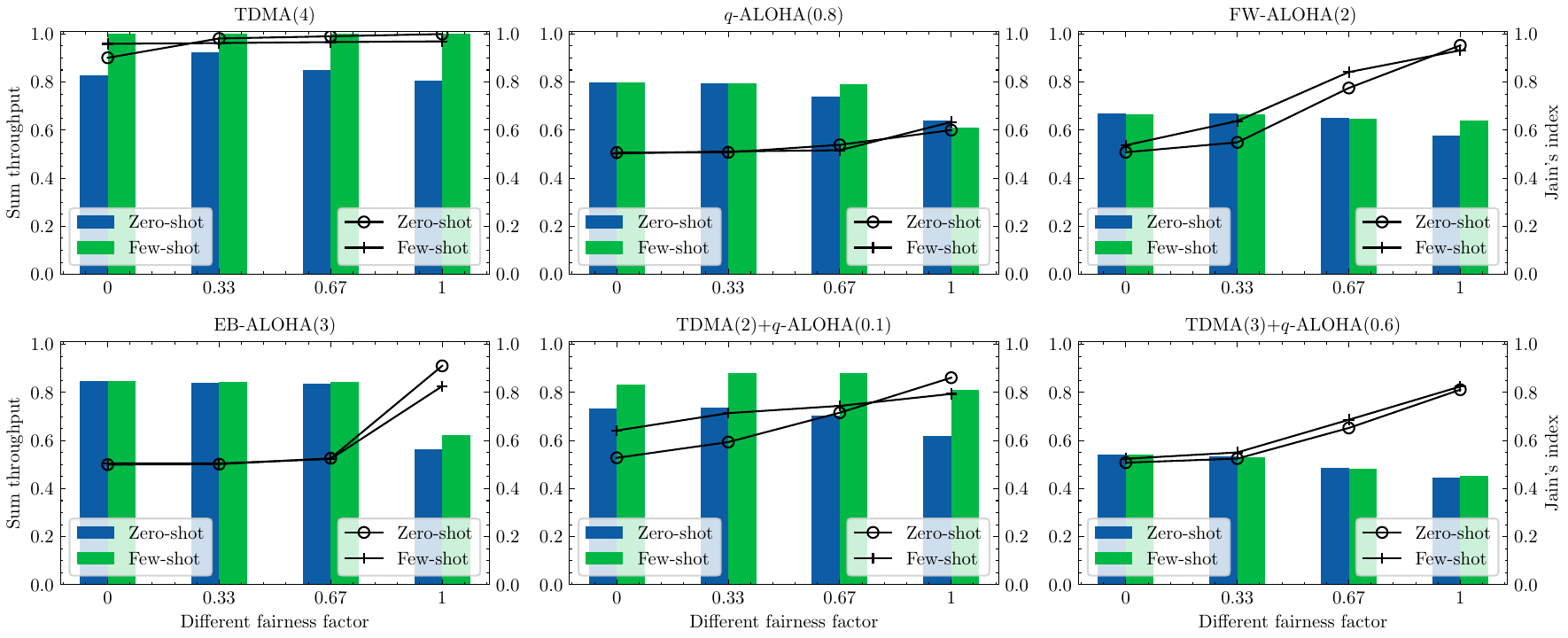}

\caption{Meta-testing with different fairness factor $\nu$. In these figures,
the bars represent the sum throughput, while the lines indicate the
Jain's index.}
\label{fig:fairness-test}
\end{figure*}

In previous subsections, our primary focus was on maximizing total
throughput. However, we recognized a potential issue with this criterion,
as certain nodes may face difficulties accessing the network in specific
environments.  To address this concern, we evaluate the reward function
with fairness consideration.  In this subsection, we maintained the
same meta-training and meta-testing sets, as well as the hyperparameters,
except for the fairness factor $\nu$ as described in Section~\ref{subsec:meta-training}.
The results presented in Fig.~\ref{fig:fairness-train} and Fig.~\ref{fig:fairness-test}
demonstrate the relationship between the fairness factor $\nu$ and
the throughput. As $\nu$ increases, the throughput of the \textquotedbl disadvantaged\textquotedbl{}
nodes exhibits a noticeable improvement, indicating that our proposed
approach effectively enhances fairness among the agent node and existing
nodes.

As shown in Fig.~\ref{fig:fairness-train}, when considering static
scheduling environments such as TDMA, increasing the value of $\nu$
does not have a significant impact on the actual access strategies.
This observation can be attributed to the fact that, in static environments,
collisions are the primary factor contributing to  the Jain's index.
However, under our specified reward scheme, collisions do not result
in higher cumulative rewards. This aligns with the practical consideration
that introducing unnecessary collisions to ensure fairness is not
meaningful. Therefore, the reward design we propose is consistent
with the requirements of static scheduling environments. However,
in environments with certain randomness and opportunistic access,
we observe that as $\nu$ increases, the throughput of nodes that
were previously unable to access in the scenario of maximizing throughput
gradually improves, indicating that they are given the opportunity
to access. Moreover, from the graph, we can see that in some environments,
increasing fairness among nodes can lead to an increase in overall
throughput. For certain environments, an appropriate fairness metric
allows the agent to overcome local optima issues and approach the
optimal strategy more effectively. Integrating appropriate fairness
factors $\nu$ during training is more conducive to achieving more
reasonable access in various environments.

Similarly, as shown in Fig.~\ref{fig:fairness-test},  in certain
environments, a reasonable fairness factor allows the agent to trade
off a slight decrease in throughput for a significant increase in
fairness, as represented by the Jain's index.

\section{Conclusion\label{sec:Conclusion}}

In this study, we proposed a generalizable MAC protocol utilizing
meta-RL to tackle the challenge of generalizing multiple access in
various heterogeneous wireless networks. Specifically, we introduced
a novel meta-RL approach that incorporates a MoE architecture  into
the representation learning of the encoder. By combining the MoE-enhanced
encoder and SAC techniques, the proposed GMA protocol effectively
learns task information by capturing latent context from recent experiences,
enabling rapid adaptation to new environments. This protocol hence
holds great promise for enhancing spectrum efficiency and achieving
efficient coexistence in heterogeneous wireless networks. Through
extensive simulations, we demonstrated that the GMA protocol achieves
universal access in training environments, with only a slight performance
loss compared to baseline methods specifically trained for each environment.
Furthermore, the GMA protocol exhibits faster convergence and higher
performance in new environments compared to baseline methods. These
results highlight the GMA protocol's capability to dynamically adapt
to different network scenarios and optimize spectrum utilization.

\appendices{}

\bibliographystyle{IEEEtran}
\bibliography{references}

\end{document}